\documentclass[12pt]{article}
\usepackage[latin1]{inputenc}
\usepackage{makeidx}
\usepackage{amsfonts}
\usepackage{graphicx}
\usepackage{amssymb}
\usepackage{amsmath}
\usepackage{subfigure}
\usepackage[margin=9pt,font=small,labelfont=bf,textfont=it,format=hang,labelformat=simple]{caption}
\begin{document}
\title{Effect of dipole-dipole interaction in two atom systems : A steady state analysis }
\author{Shaik Ahmed, Ashoka Vudayagiri and P Anantha Lakshmi \footnote{ corresponding author: palsp@uohyd.ernet.in}\\School of Physics, University of Hyderabad\\ Hyderabad - 500046, India}
%\ead{palsp@uohyd.ernet.in}
\maketitle

\begin{abstract}
The behaviour of a system of two two-level atoms, both identical as well as non-identical, under the influence of a continuously varying dipole-dipole coupling parameter for a wide range of other parameters such as the atomic level separations and the coupling strength of the external radiation field is explored.  A detailed analysis of the behaviour of various level populations and some of the atomic coherences is presented.  The influence of the competing effects  between the Rabi field strength and the dipole dipole interaction on the two photon absorption probability is explored in detail both for identical as well as non-identical atoms. There are significant variations in the behaviour of identical and non-identical atoms as a function of the dipole coupling strength, even for small amounts of non-identity of the atoms, which in this study is incorporated through the atomic level separations. 

\end{abstract}%----------------------------------------------------

\section{INTRODUCTION} 

Cooperative effects \cite{r1}, for example collective emission, are a general characteristic of macroscopic systems consisting of a large number of atoms confined in a region of space much smaller than an optical wavelength.  However, in order to understand the collective behaviour of macroscopic systems, it is essential that we have a microscopic formulation of the interaction between the atoms and the electrons.  
Mention has to be made that cooperative effects are not important in most experiments involving interaction of atoms with laser fields, which are in the dilute gas regime, where the confinement region is much larger than an optical wavelength.  However, situations such as in a Rydberg gas where the dipole moments are very large or in Bose Einstein Condensates (BEC) where the densities are higher, the collective behaviour of atoms indeed has strong influence on the outcome of the experiment.  

We study a prototype model consisting of a pair of interacting atoms in the presence of a radiation field of arbitrary strength.  Even though such a model may seem rather elementary, the study is significant since this model can  serve as a basic building block in understanding of the cooperative effects among interacting atoms.  Recent advances have enabled trapping and studying of very few  atoms in a trap, in light of which this  study is all the more relevant.

A large amount of work exists that study cooperative effects among few atoms \cite{r1,r2,r3,r4,r5,r6,r7,r8,r9,r10}, of  particular interest being the case of a single pair of two-level atoms.  The justification for the latter comes from the fact that collective phenomena basically are a result of pair-wise interaction among atoms and hence the study of two atoms forms the  basic building block in treating the collective interactions. One such earlier work \cite{r11} reports that correlation between two identical atoms can give rise to new resonances in the resultant fluorescence spectrum.  In  another study\cite{r12} it has been shown that dipole-dipole interaction between atoms may lead to the suppression of the fluorescence whenever the atoms are placed within one optical wavelength.  In reference \cite{r13}, Kim et al. have shown that cooperative effects can cause modification in transition rates and level shifts, thus giving rise to significant changes in the multiple jump dynamics.

In yet another study, Varada and Agarwal \cite{r14} have observed an enhancement of the two photon resonance due to the dipole-dipole interaction strengths.  Another aspect, namely, the entanglement characteristics of such  systems were explored via a study of two photon correlations by von Zanthier et al.\cite{r15}.  

In general, in a collection of atoms, it is observed that, the first excited atom prevents the excitation of neighbouring atoms due to the dipole-dipole interaction, which is referred to as the dipole blockade\cite{r16,r17}. 
 This dipole blockade can give rise to the production of single atom excited collective states.
 
 In a more recent work \cite{r18} the effect of dipole blockade on the dynamical evolution of two two-level atom system in the presence of a driving field was studied to  understand  the entanglement characteristics of the system. 
Rydberg atoms, by virtue of their large dipole moments exhibit strong dipole-dipole interactions which can give rise to entanglement \cite{r19,r20,r21}.  This interesting property of Rydberg atoms has led to proposals for using them in quantum computation schemes \cite{r22}.

 In view of the rich and varied applications the system of two two-level atoms has to offer, it becomes an interesting subject of study.  One ought to mention that most of the earlier work reported is carried out in the dressed state formalism.  In this paper, we examine in detail, the behaviour of the two two-level atom system both for identical as well as non-identical atoms.  We work in the product space of the single atom states as we believe that scaling up to more atoms is straight forward.  In this basis, the two two-level atom system can be modeled as an effective four level atom system.   

The study carried out here is valid for arbitrary strengths of various interactions namely the electric dipole interaction between the atom and the external radiation field, the dipole-dipole interaction between atoms.  The master equation approach \cite{r23} is used to incorporate the spontaneous emission decays.

Density matrix formalism is employed to obtain the equation of motion for the various atomic level populations and coherences.  The resulting sixteen coupled differential equations governing the dynamics of the system can be solved both in steady state regime as well as in time domain.  Emphasis is laid on exploring the influence of competing behaviour  between  the Rabi field strength and the dipole interaction on the level populations, atomic coherences and the total one photon as well as two photon absorption cross sections, across different frequency regimes. 

As mentioned earlier, Varada and Agarwal\cite{r14} have reported the enhancement of two-photon resonance due to the presence of dipole-dipole coupling between non-identical atoms.   In this paper, we have extended their study to a wider parameter regime to explore the behaviour of this two-photon resonance in greater detail.  We have observed that the enhancement of the two-photon resonance is not a monotonic feature.  Instead, beyond a certain critical value of the dipole coupling strength, it begins to decrease again.  It is this feature of the two-photon resonance that has been addressed in detail in this paper.

The organization of this paper is as follows. In Sec. II, we present the mathematical formulation of the problem.  The  basic equations that govern the dynamics of the system of two two-level atoms interacting with the single-mode laser field are written down in the rotating wave approximation.  In section III, numerical results in steady state of various quantities are presented for a wide choice of parameters, both for identical as well as non-identical atoms.   It is seen that a small deviation from identity of the atoms gives rise to significant changes in the behaviour of the system.

\section{THE MODEL}
\begin{figure}[h!]
\vspace{-15pt}
\centering
\includegraphics[height=8 cm,width=10 cm]{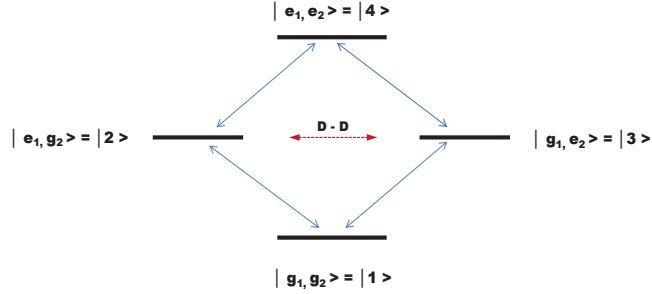} 
\vspace{-70pt}
\caption{Schematic energy level diagram of two two-level atoms as an equivalent four-level system}
\label{fig:One}
\end{figure}

We consider two isolated two level atoms fixed in position with a separation 
 \textbf{r}, driven by a single mode resonant laser field, of frequency $\omega_L$ (wavelength $\lambda$), which is nearly resonant with the transition frequency of one of the atoms, in the region $\textbf{r}\ll\lambda$.  Here $|\textbf{g}_{i}\rangle$ and $|\textbf{e}_{i}\rangle$ are respectively the ground and excited states of the $i^{th}$ atom ( i=1,2).  The interaction with the field induces  dipole moment in the atoms, which in turn interact with each other via photon exchange.
 In this model we assume that the external field is propagating perpendicular to the inter atomic axis so  that the interaction of  field with the atoms is purely symmetric.  We use the semi-classical approach to study the problem in which the atomic system is treated quantum mechanically while the laser field interaction as well as the dipole - dipole interaction are treated classically.  As mentioned earlier, the system of two two-level atoms  can be represented as an effective four level system as shown in figure \ref{fig:One}.

The product state $|g_{1}g_{2}\rangle$ ($|e_{1}e_{2}\rangle$) corresponds to the situation where both atoms are in their ground(excited)state.  The intermediate state $|g_{1}e_{2}\rangle$ ($|e_{1}g_{2}\rangle$) represents the case when the first (second) atom is in ground(excited) state and second(first) atom is in excited(ground) state.  For brevity the following notation is introduced.  $|g_{1}g_{2}\rangle =|\textbf{1}\rangle,|e_{1}g_{2}\rangle =|\textbf{2}\rangle,|g_{1}e_{2}\rangle =|\textbf{3}\rangle$ and $|e_{1}e_{2}\rangle =|\textbf{4}\rangle$.
 The Hamiltonian for this two two-level atomic  system can be written as

 \begin{eqnarray}
 H=\begin{bmatrix}
\omega_{1} & -\alpha^{*} _{21} & -\alpha^{*} _{31} & 0 \\
-\alpha _{21} & \omega_{2} & -g & -\alpha^{*} _{42} \\
-\alpha_{31} & -g & \omega_{3} & -\alpha^{*} _{43} \\
0 & -\alpha _{42} & -\alpha _{43}& \omega_{4}\end{bmatrix} 
\end{eqnarray}
 
where ${\omega_{i}}$(i=1,2,3,4) is the frequency of the $ \textbf{i}^{th}$ level, $\alpha_{ij}$ are the different Rabi frequencies which represent different atom-field coupling strengths and $ g $ represents the strength of the dipole - dipole interaction.  

Several theoretical approaches can be used to treat the system of two atoms interacting with a quantized radiation field.  One of the more popular approaches is the master equation method to obtain the equation of motion for the density operator of the atomic system coupled to a reservoir (the multimode vacuum field).  The master equation approach enables one to treat the evolution of the atom plus field system entirely in terms of the atomic operators.  In this work, we use the master equation formalism to study the dynamical evolution of the density operator of the atomic system.
The equation of motion for the density operator $\rho$ is given by

\begin{equation}
i\hbar\frac{\partial\rho}{\partial t}=[H,\rho]-\mathcal{L} \rho
\end{equation}

 The first term on the right hand side  represents the interaction with the radiation field and the dipole-dipole  interaction whereas the second term (The Liouvillean) accounts for different decay mechanisms.  Mention has to be made that in the equations of motion that are presented in the following, the spontaneous emission decays alone have been included.  However, the equations are suitably modified to take into account the cooperative decay wherever in the calculations, the role of the cooperative decay parameter also has been examined by suitable modification of the equations of motion.  The resulting equations of motion for the sixteen density matrix elements in the rotating wave approximation are obtained as follows:
{\begin{footnotesize}\begin{align}
&{\partial_t\tilde{\rho_{11}}}=i\alpha^{*}_{21}\tilde{\rho_{21}} - i\alpha_{21}\tilde{\rho_{12}}+i\alpha^{*}_{31}\tilde{\rho_{31}} -i\alpha_{31}\tilde{\rho_{13}}+2\gamma_{21}\tilde{\rho_{22}}+2\gamma_{31}\tilde{\rho_{33}}\nonumber\\
&{\partial_t\tilde{\rho_{22}}} = -i\alpha^{*}_{21}\tilde{\rho_{21}}+i\alpha_{21}\tilde{\rho_{12}}+ig(\tilde{\rho_{32}}-\tilde{\rho_{23}})+i\alpha^{*}_{42}\tilde{\rho_{42}}-i\alpha_{42}\tilde{\rho_{24}}+2\gamma_{42}\tilde{\rho_{44}}-2\gamma_{21}\tilde{\rho_{22}}\nonumber\\
&{\partial_t\tilde{\rho_{33}}} = -i\alpha^{*}_{31}\tilde{\rho_{31}}+i\alpha_{31}\tilde{\rho_{13}}-ig(\tilde{\rho_{32}}-\tilde{\rho_{23}})+i\alpha^{*}_{43}\tilde{\rho_{43}}-i\alpha_{43}\tilde{\rho_{34}}+2\gamma_{43}\tilde{\rho_{44}}-2\gamma_{31}\tilde{\rho_{33}}\nonumber\\
&{\partial_t\tilde{\rho_{44}}} = i\alpha_{42}\tilde{\rho_{24}}+i\alpha_{43}\tilde{\rho_{34}}-i\alpha^{*}_{42}\tilde{\rho_{42}}-i\alpha^{*}_{43}\tilde{\rho_{43}}-(2\gamma_{42}+2\gamma_{43})\tilde{\rho_{44}}\nonumber\\
&{\partial_t\tilde{\rho_{12}}} = (i\Delta_{1}-\gamma_{21})\tilde{\rho_{12}}-i\alpha^{*}_{21}\tilde{\rho_{11}}+i\alpha^{*}_{21}\tilde{\rho_{22}}+i\alpha^{*}_{31}\tilde{\rho_{32}}-ig\tilde{\rho_{13}}-i\alpha_{42}\tilde{\rho_{14}}\\
&{\partial_t\tilde{\rho_{13}}} =(i\Delta_{2}-\gamma_{31})\tilde{\rho_{13}}-i\alpha^{*}_{31}\tilde{\rho_{11}}+i\alpha^{*}_{31}\tilde{\rho_{33}}+i\alpha^{*}_{21}\tilde{\rho_{23}}-ig\tilde{\rho_{12}}-i\alpha_{43}\tilde{\rho_{14}}\nonumber\\
&{\partial_t\tilde{\rho_{14}}} =(i\Delta_{1}+i\Delta_{2}-\gamma_{43}-\gamma_{42})\tilde{\rho_{14}}-i\alpha^{*}_{42}\tilde{\rho_{12}}-i\alpha^{*}_{43}\tilde{\rho_{13}}+i\alpha^{*}_{21}\tilde{\rho_{24}}+i\alpha^{*}_{31}\tilde{\rho_{34}}\nonumber\\
&{\partial_t\tilde{\rho_{23}}}=(i\Delta_{2}-i\Delta_{1}-\gamma_{21}-\gamma_{31})\tilde{\rho_{23}}+i\alpha_{21}\tilde{\rho_{13}}-i\alpha^{*}_{31}\tilde{\rho_{21}}+i\alpha^{*}_{42}\tilde{\rho_{43}}-i\alpha_{43}\tilde{\rho_{24}}+ig(\tilde{\rho_{33}}-\tilde{\rho_{22}})\nonumber\\
&{\partial_t\tilde{\rho_{24}}} =(i\Delta_{2}-\gamma_{21}-\gamma_{42}-\gamma_{43})\tilde{\rho_{24}}+i\alpha_{21}\tilde{\rho_{14}}+i\alpha^{*}_{42}\tilde{\rho_{44}}-i\alpha^{*}_{42}\tilde{\rho_{22}}-i\alpha_{43}\tilde{\rho_{23}}+ig\tilde{\rho_{34}}\nonumber\\
&{\partial_t\tilde{\rho_{34}}} =(i\Delta_{1}-\gamma_{31}-\gamma_{42}-\gamma_{43})\tilde{\rho_{34}}+i\alpha_{31}\tilde{\rho_{14}}+i\alpha^{*}_{43}\tilde{\rho_{44}}-i\alpha^{*}_{43}\tilde{\rho_{33}}-i\alpha_{42}\tilde{\rho_{32}}+ig\tilde{\rho_{24}}\nonumber
\end{align}\end{footnotesize}}
The remaining equations can be obtained by realizing that 
\begin{align}
&{\partial_t\tilde{\rho_{ij}}} = {\partial_t\tilde{\rho_{ji}}^{*}} . \nonumber
\end{align}
In the above ~ $\Delta_{1}=\omega_{atom1}-\omega_L, \qquad \Delta_{2}=\omega_{atom2}-\omega_L,$  and the total detuning~ $\Delta=\Delta_{1}+\Delta_{2}.$ \\  Here $2\gamma_{ij}$ represents the rate of spontaneous decay from level $j$ to $i$ while  $\gamma_{ij}$ is the decoherence rate.  By using the completeness condition $\rho_{11}+\rho_{22}+\rho_{33}+\rho_{44}=1 $, the 16 equations can be reduced to 15 equations which can be cast in the form 

\begin{equation}
\frac{\partial\Psi}{\partial t}=M\Psi+\Phi
\end{equation} 
where $M$ is a $ 15 \times 15$  coefficient matrix  and   $\Psi$ and $\Phi$ are column vectors  each of length 15 which are defined in the following.

\begin{align} 
 &\Psi  = [ \rho_{11}~  \rho_{12}~  \rho_{13}~  \rho_{14}~  \rho_{21}~  \rho_{22}~  \rho_{23}~  \rho_{24}~  \rho_{31}~  \rho_{32}~  \rho_{33}~  \rho_{34}~  \rho_{41}~  \rho_{42}~  \rho_{43}]^T \\
 &\Phi   = [ 0~ 0~ 0~ 0~ 0~  2\gamma_{42}~ 0~ i \alpha^{*}_{42}~ 0~ 0~ 2\gamma_{43}~ i\alpha^{*}_{43}~ 0~ -i  \alpha_{42}~ -i\alpha_{43}]^T .
 \end{align}

The steady state solution of the density matrix elements may be  obtained as
\begin{equation}
\Psi_{ss} = \Psi(t\to\infty)=-M^{-1}\Phi
\end{equation}
 
 In the next section we present some of our numerical results of various level populations and coherences in the steady state, for both identical as well as non-identical atoms.  With no loss of generality, we assume that the coupling between the atom and the field to be same for both atoms, which corresponds to all $\alpha_{ij}$'s being equal which is denoted by $\alpha$ and all the spontaneous decay rates also are taken to be equal($\gamma_{ij}=\gamma$)
 
\section{RESULTS AND DISCUSSION}
The steady state behaviour of level populations, various atomic coherences and absorption cross sections  has been studied numerically for a wide range of parameter regimes, both for identical and non-identical atoms.  In this section, some representative numerical results for typical values of the parameters are being presented.

 \subsection{Identical atoms}
\begin{figure}[h!]
\centering
\subfigure[]{
\includegraphics[width=6.5 cm,height=6 cm]{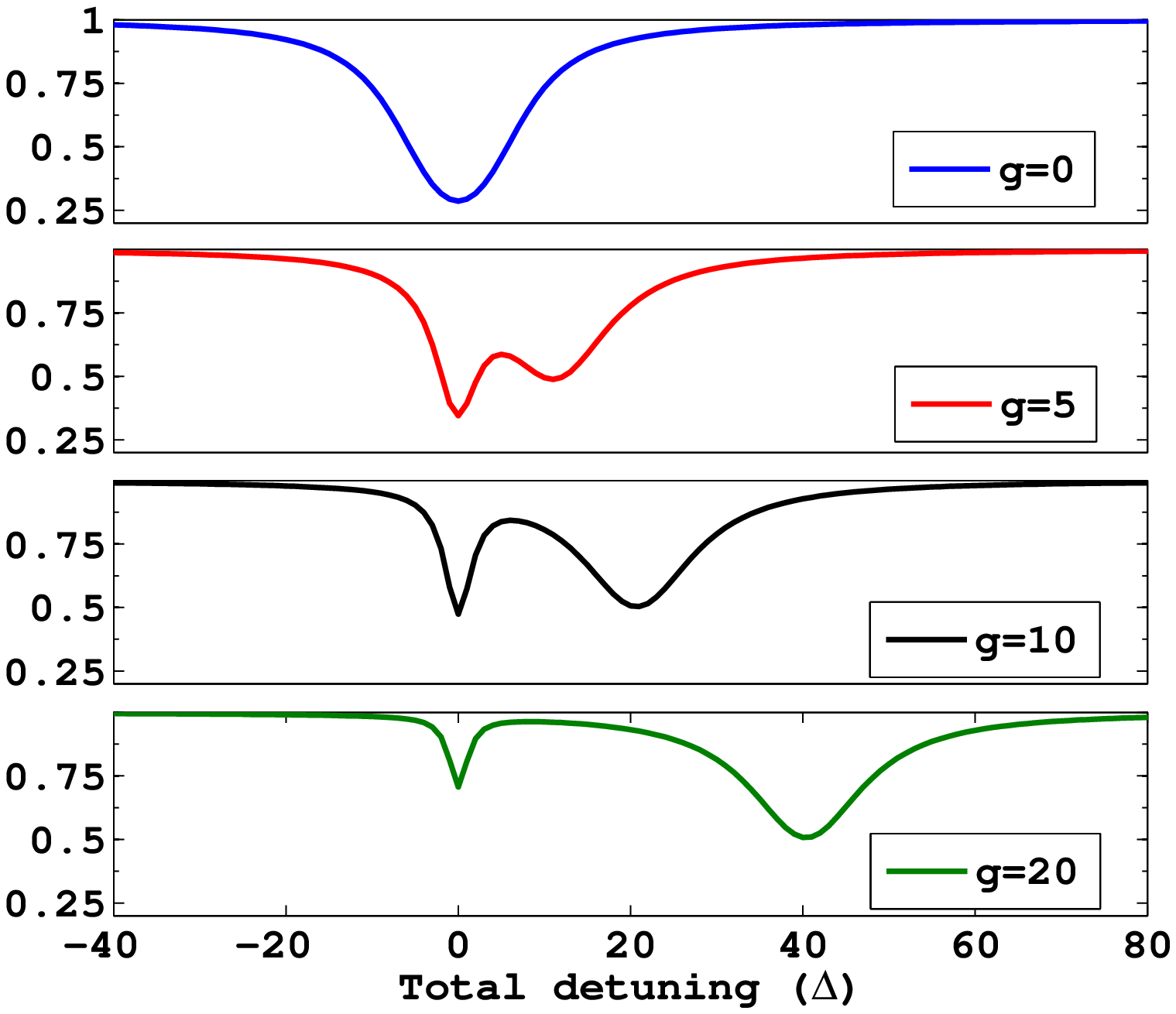}
\label{fig:subfig1}
}
\subfigure[]{
\includegraphics[width=6.5 cm,height=6 cm]{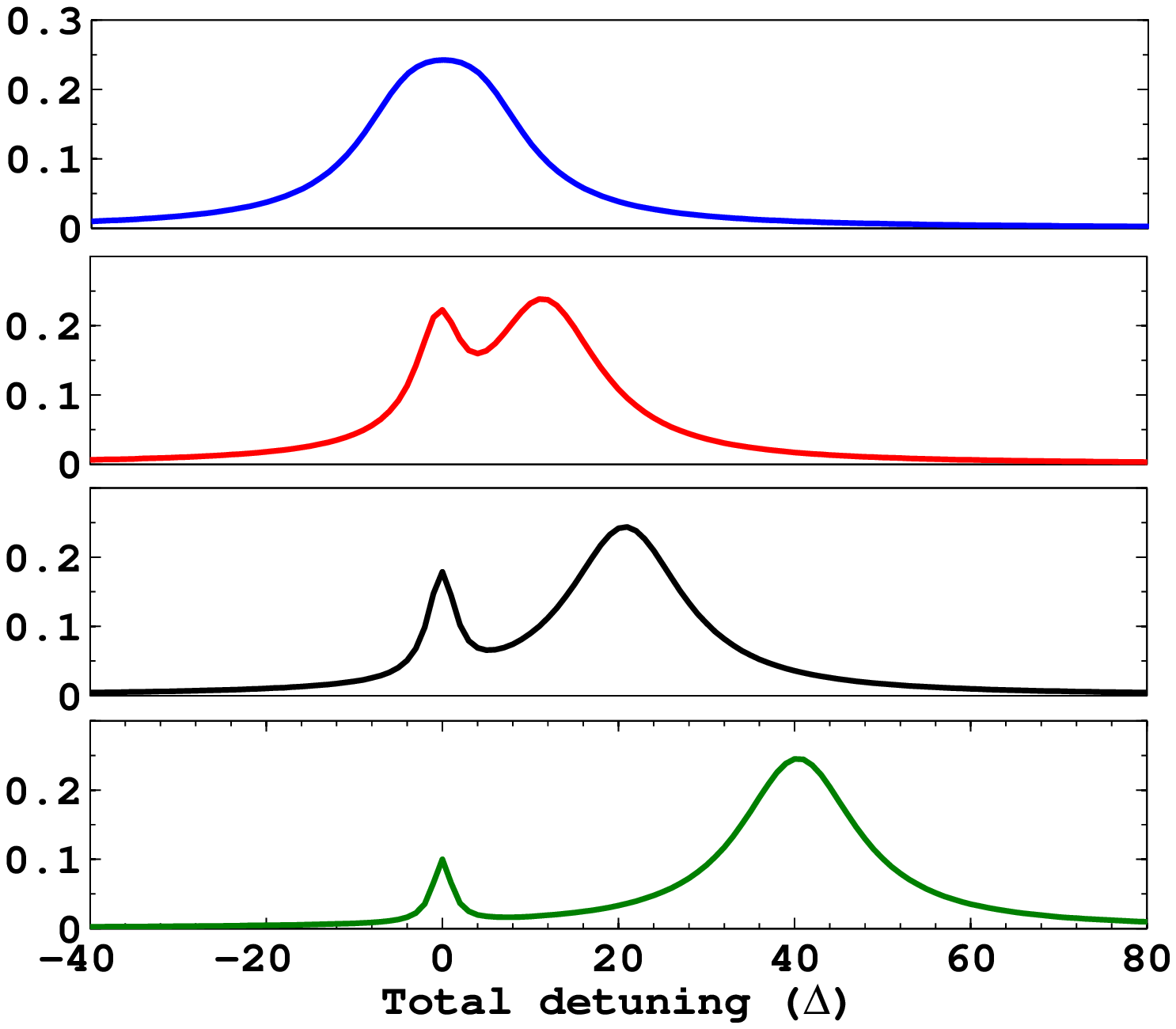}
\label{fig:subfig2}
}
\subfigure[]{
\includegraphics[width=6.5 cm,height=6 cm]{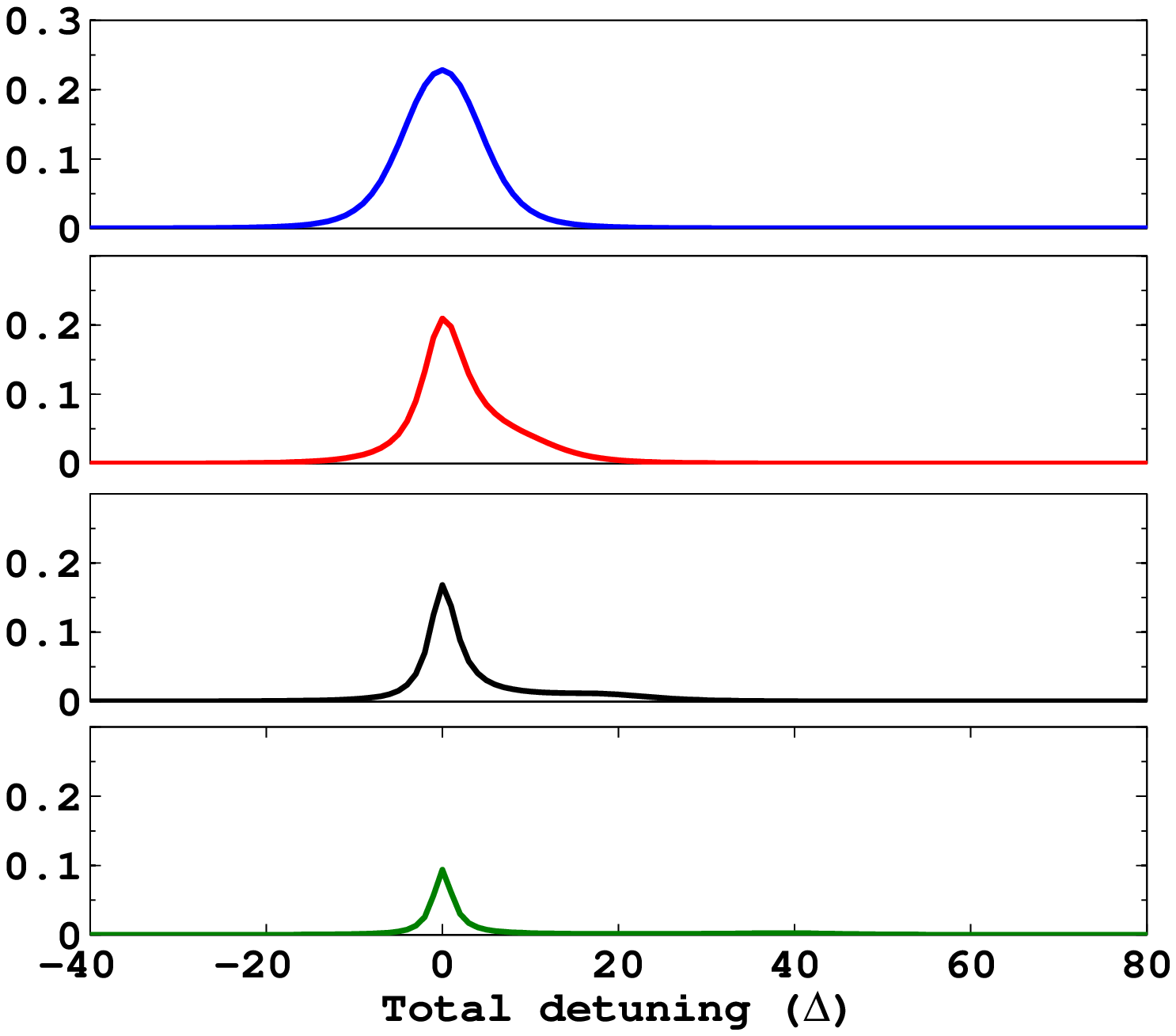}
\label{fig:subfig3}
}
\subfigure[]{
\includegraphics[width=6.5 cm,height=6 cm]{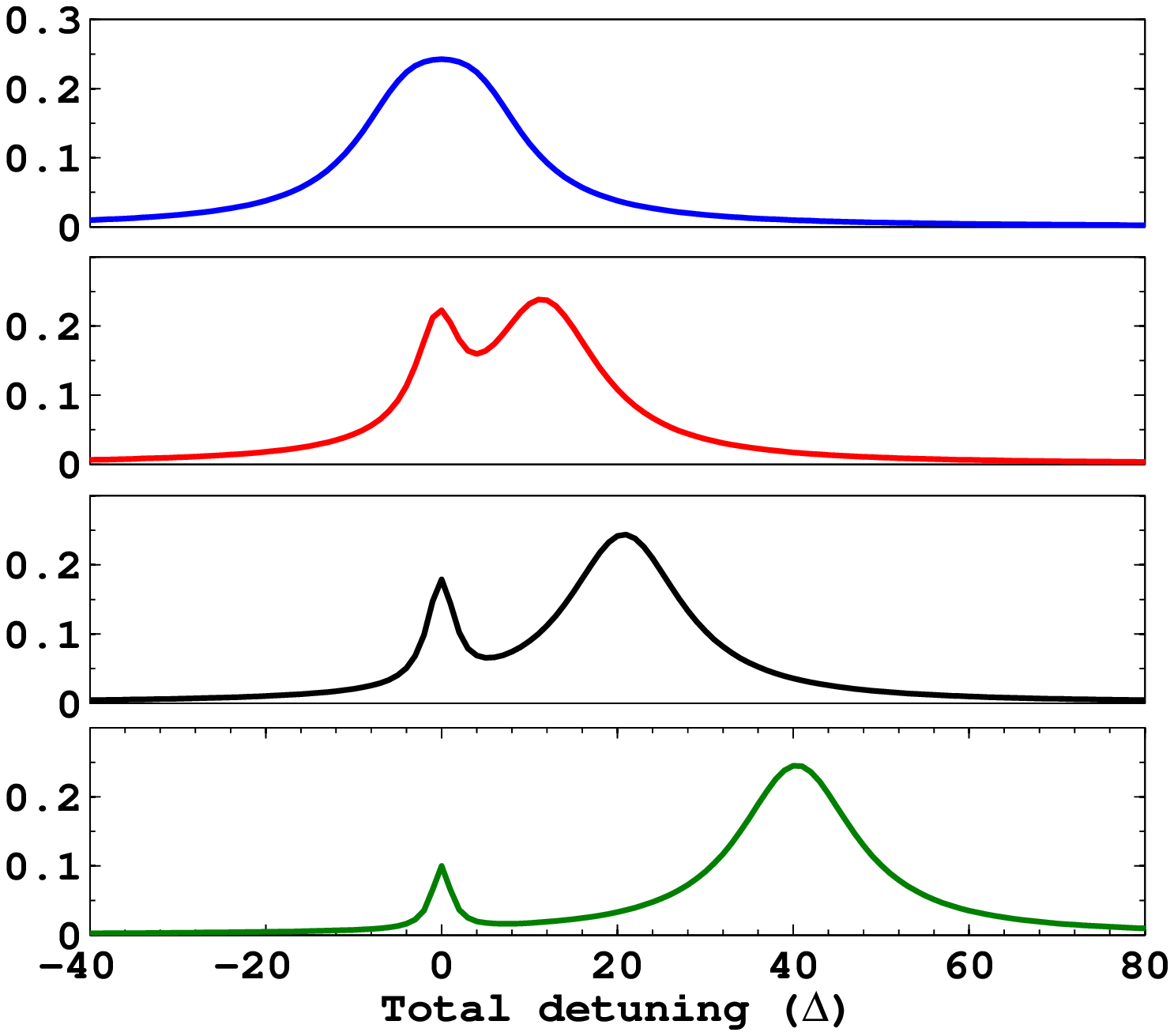}
\label{fig:subfig4}
}
\caption[Optional caption for list of figures]{\textit{Level populations for identical atoms}  \subref{fig:subfig1} $\rho_{11}$, \subref{fig:subfig2} $\rho_{22}$, \subref{fig:subfig3} $\rho_{44}$ and \subref{fig:subfig4} $\rho_{33}$}
\label{fig:Two}
\end{figure}

The steady state populations $\rho_{ii},i=1,2,3,4$ for identical atoms, are plotted as a function of the total detuning $\Delta = \Delta_1 + \Delta_2$, in figure \ref{fig:Two} for a given value of the Rabi frequency and for different values of the dipole coupling parameter $ g $.  Sub-figures \ref{fig:subfig1}, \ref{fig:subfig2}, \ref{fig:subfig3} and \ref{fig:subfig4} respectively show $\rho_{11}, \rho_{22}, \rho_{44}$ and $\rho_{33}$.  Each of the sub-figures corresponds to different values of the dipole coupling parameter namely $ g=0, 5, 10$ and $20$ from top to bottom.  As already mentioned, the value of Rabi frequency is held constant for all of them at $\alpha = 2$.  It is to be noted that all parameters for all the data presented in this paper are normalized to $2\gamma$, the population decay rate.  

When there are no dipole-dipole interactions, i.e., $g=0$, the two atoms are independent of each other and hence show single resonant peak at $\Delta=0$, characteristic of a single two-level atom (topmost curves in all the four subfigures \ref{fig:subfig1}-\ref{fig:subfig4}). 

All possible configurations, such as both atoms being in their respective ground state ($\rho_{11}$), any one of them in the excited state while the other in ground state ($\rho_{22}$ and $\rho_{33}$) and both atoms in excited state ($\rho_{44}$), are all equally probable and hence show a dip/peak value of 0.25 at resonance.  The resonant dip in $\rho_{11}$ is matched equally by the increase in $\rho_{22}, \rho_{33}$ and $\rho_{44}$.  The rest of the curves in figure\ref{fig:Two} correspond to non zero values of $ g  $ in increasing order from top to bottom as indicated in the figure.  The salient features observed in these curves are as follows: (i) The central resonant peak decreases in height(depth in case of $\rho_{11}$) with increase in $ g $. (ii) A new side band appears at a non zero detuning $\Delta$.  Though it is very difficult to solve the equations analytically, we have nevertheless been able to deduce, from the numerical values computed over a wide range of parameter values, that this peak appears approximately at $\Delta = 2\sqrt{(|g|^2+2\alpha^2)}$. (iii) This side band does not exist for $\rho_{44}$ ( \ref{fig:subfig3}), although a small protuberance can be seen for intermediate values of g. (iv) The populations $\rho_{22}$ and $\rho_{33}$ are identical, as expected, indicating that single atom excitation does not distinguish between the two atoms. (v) With increasing $ g $, the value of the resonant peak/dip  at $\Delta = 0$ identically decreases for all the four states.  This re-affirms that (a) the energy levels are shifted due to the dipole-dipole interaction \cite{r24} and (b) the dipole-dipole interaction causes a blockade, due to which the probability for both atoms to be  simultaneously excited ($\rho_{44}$) reduces drastically.
\subsection{Non-identical atoms}
For the case of non-identical atoms, we assume different energy level separation between ground and excited states of the two atoms and hence $\Delta_1\ne\Delta_2$.  We also assume that the two decay constants  $\gamma_{21}$ and $\gamma_{31}$ are close enough to be assumed equal  which we denote by  $\gamma$.  The Rabi frequencies $\alpha_{21}$ and $\alpha_{31}$ are also taken to be equal to $\alpha$.  Without a need to choose specific atomic species, we assume a situation of $\Delta_2= \Delta_1+10$, in units of $2\gamma$, the population decay rate\footnote{The two atoms need not be very different from each other.  Instead of two different species, one could always use the same atoms, but two different transitions based on different polarizations and yet achieve the same effect.  Although it is almost impossible to obtain two atoms such that their decay rates as well as dipole transition elements are exactly identical, one can choose a situation to begin with wherein the differences are insignificant compared to the energy level separation thus justifying our choice of various parameters above \cite{r25}.}.
 \begin{figure}[h!]
\centering
\subfigure[]{
\includegraphics[width=6.5 cm,height=6 cm]{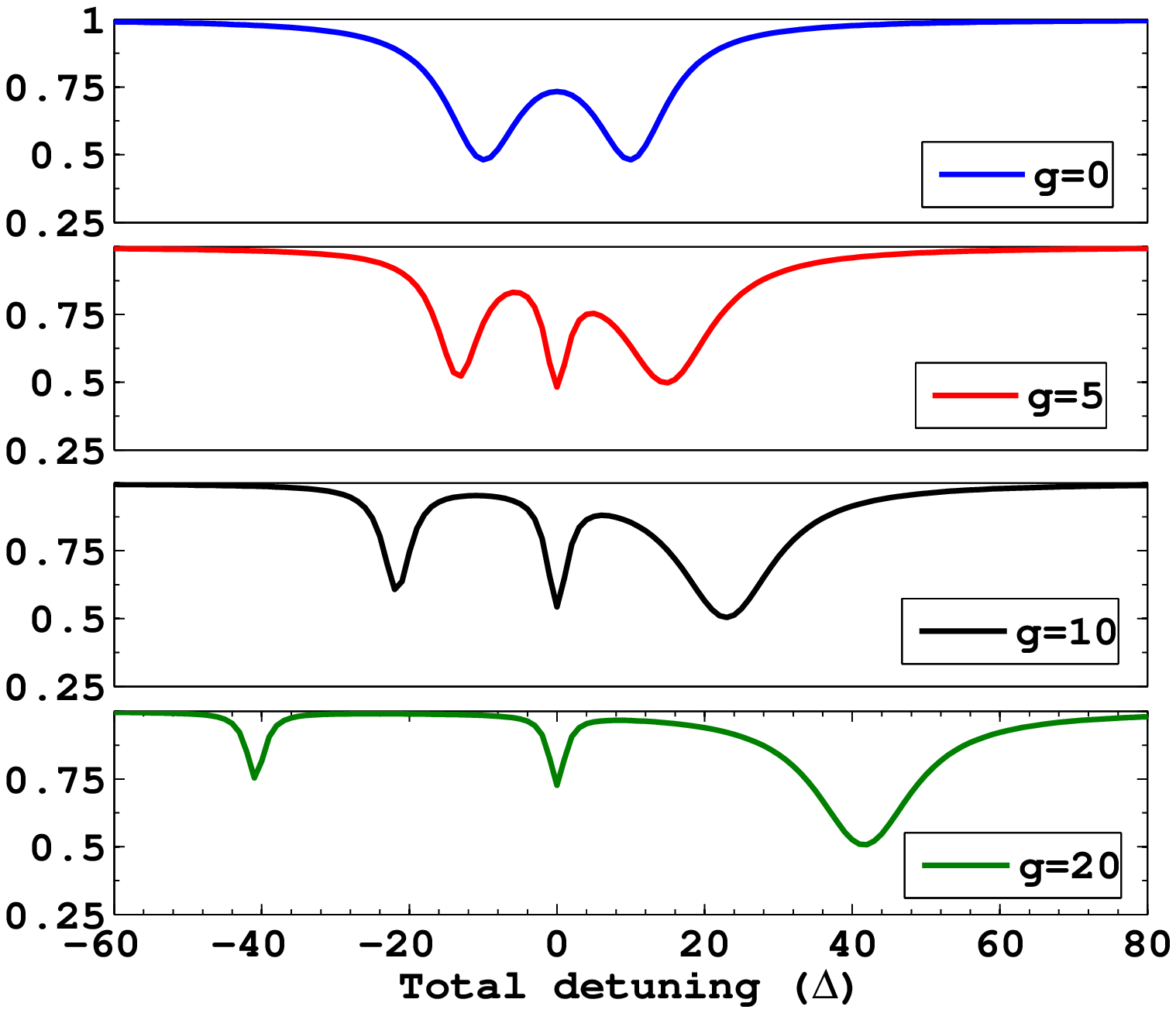}
\label{fig:subfigure1}
}
\subfigure[]{
\includegraphics[width=6.5 cm,height=6 cm]{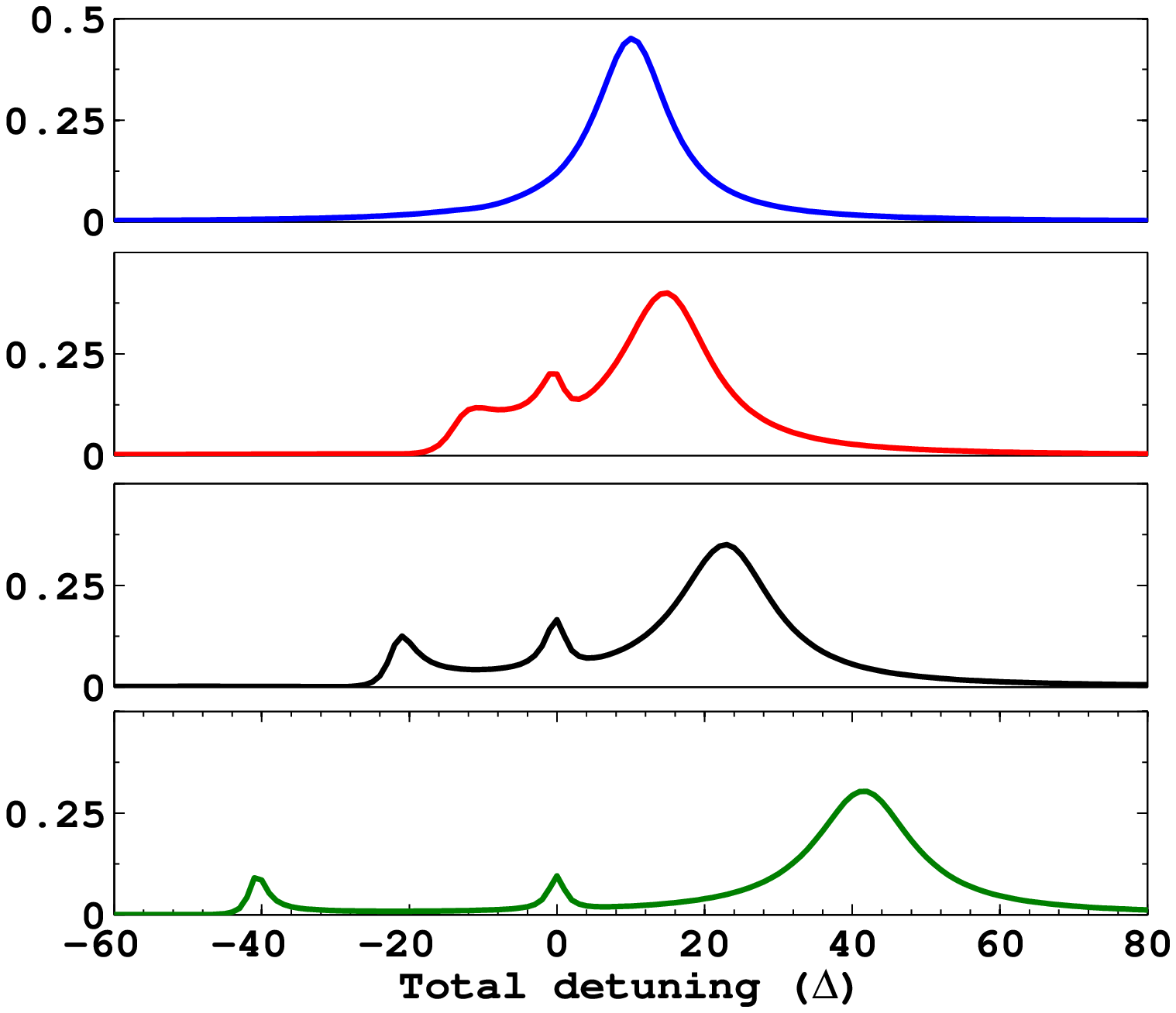}
\label{fig:subfigure2}
}
\subfigure[]{
\includegraphics[width=6.5 cm,height=6 cm]{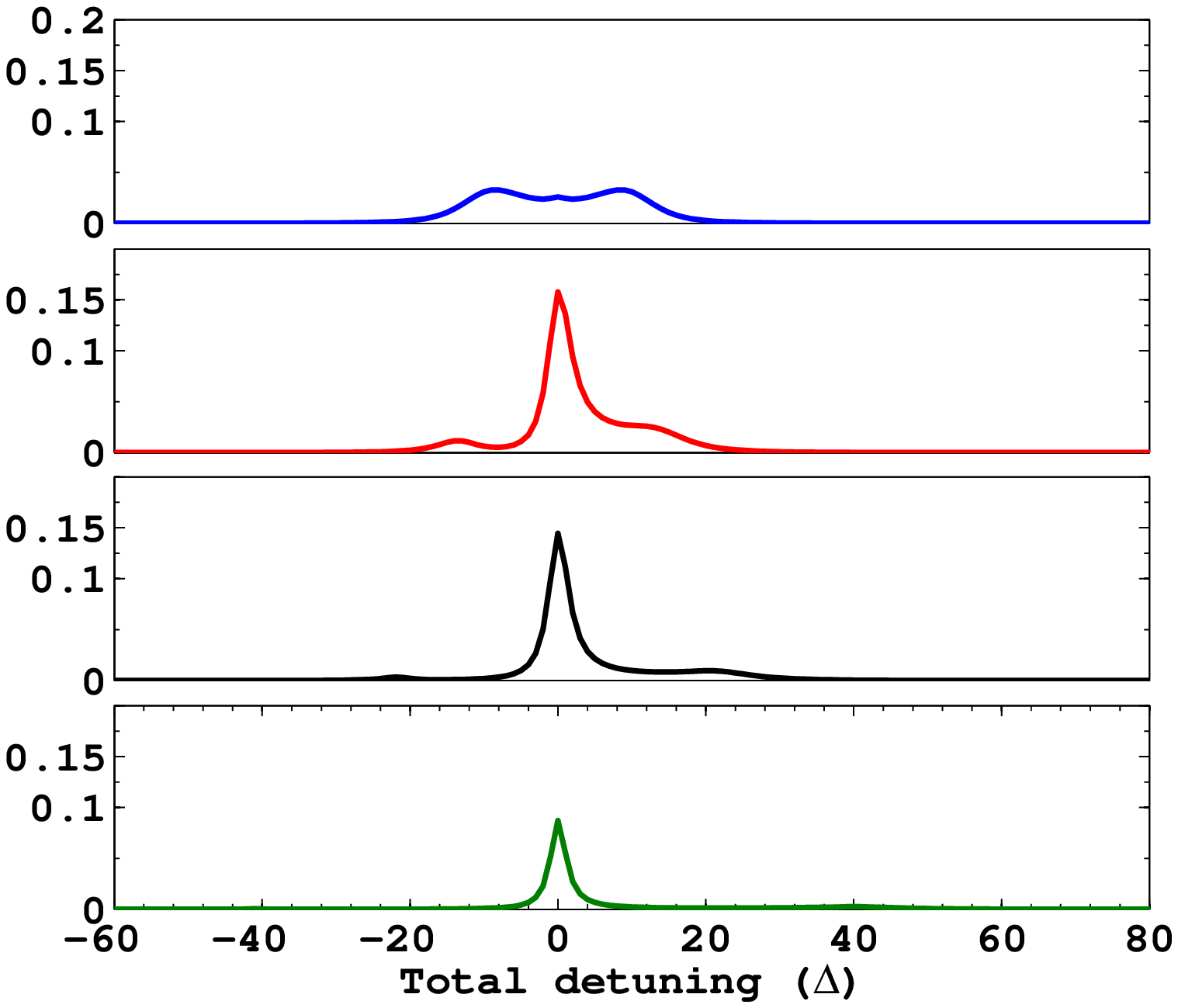}
\label{fig:subfigure3}
}
\subfigure[]{
\includegraphics[width=6.5 cm,height=6 cm]{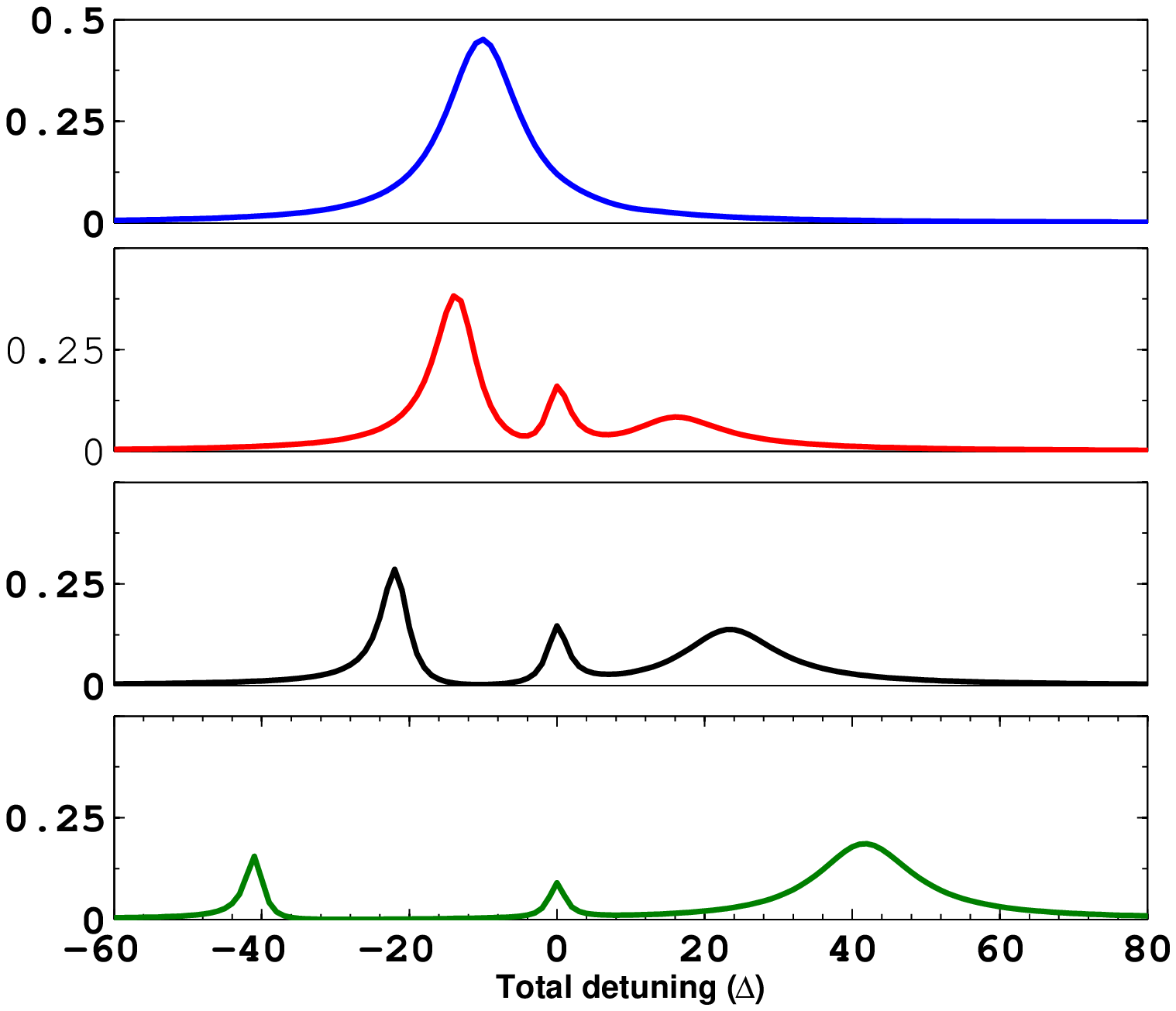}
\label{fig:subfigure4}
}
\caption[Alternative caption for list of figures]{\textit{Level populations for non-identical atoms}  \subref{fig:subfigure1} $\rho_{11}$, \subref{fig:subfigure2} $\rho_{22}$, \subref{fig:subfigure3} $\rho_{44}$ and \subref{fig:subfigure4} $\rho_{33}$}
\label{fig:Three}
\end{figure}

The steady state level populations for the case of non-identical atoms, as a function of the total detuning $\Delta = \Delta_1+\Delta_2$, are presented in figure [\ref{fig:Three}].  In case of $ g = 0 $, it can be noticed that there are two resonances (dips) for the ground state population, as seen in the top curve of figure \ref{fig:subfigure1}.  These two resonances respectively occur when  $\Delta_1=0$ and $\Delta_2=0$, and hence are on either side of the line-center $\Delta$= 0.  Corresponding to these two dips in the ground state population, a single resonance peak is seen in each of $\rho_{22}$ (at $\Delta_1 = 0$) and $\rho_{33}$(at $\Delta_2 = 0$).  $\rho_{44}$ shows two very small peaks, again at exactly the same positions as the dips  in  $\rho_{11}$.  As the dipole coupling strength $g$ is increased, side bands start to appear at $+2\sqrt{(|g|^2+\delta^2/4)+2\alpha^2}$ and $-2\sqrt{(|g|^2+\delta^2/4)-2\alpha^2}$, where $\delta = \Delta_1-\Delta_2$. For higher values of $g$, the side bands are dominant only in $\rho_{22}$ and $\rho_{33}$, corresponding to which there is hardly  any activity in $\rho_{44}$.  The two-atom excited state $\rho_{44}$, however looses the twin resonant peaks  which were present in the absence of dipole - dipole interaction $ g $, but develops a single peak at total detuning $\Delta=0$.  This is in concurrence with the result of Varada and Agarwal \cite{r14}, indicating that as the strength of dipole-dipole interaction $g$ increases, the probability for stepwise excitation of  both atoms decreases.  At the same time, the probability for  simultaneous excitation of both atoms through a two-photon absorption process increases due to the influence of $g$, as indicated by the central resonance peak of $\rho_{44}$.   Interestingly, the height of $\rho_{44}$ peak  which is appearing only in the presence of $g$ does not increase monotonically with an increase in $g$.  There is an optimum value of $g$ corresponding to a given value of the Rabi field strength $\alpha$, at which $\rho_{44}$ attains its  maximum value.  Further increase in $g$ seems to result in a decrease in the peak height of $\rho_{44}$, indicating a decrease in the probability for simultaneous absorption of two photons by  the atoms, a feature which is present only for the case of non-identical atoms.  The corresponding peaks in $\rho_{22}$ and $\rho_{33}$ at $\Delta=0$ can be interpreted as due to decay of population from $\rho_{44}$.   The fluorescence from the system which is a linear combination of the level populations, namely $\rho_{22} + \rho_{33} + 2 \rho_{44}$, shows qualitative agreement with the experiment involving two terylene molecules \cite{r25}.  We do observe an additional feature, namely the non-monotonic increase of the central peak as the dipole coupling strength is steadily increased.

\begin{figure}[h!]
\centering
\includegraphics[width=14 cm,height=8 cm]{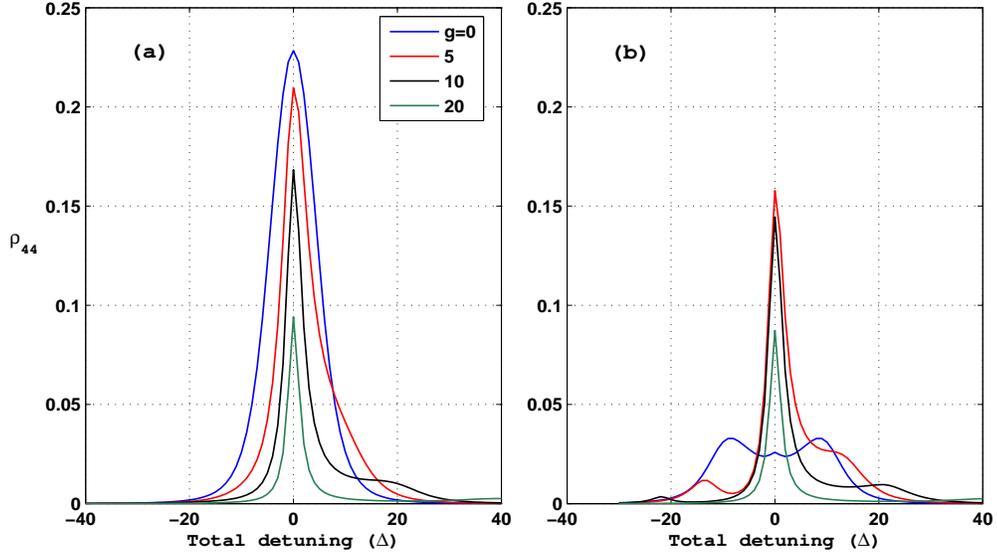} 
\caption{Level population $\rho_{44}$ for the case of both (a) identical and (b) non-identical atoms for Rabi frequency $ \alpha $ = 2 }

\label{fig:Four}
\end{figure}

\begin{figure}[h!]
\vspace{-10pt}
\centering
\subfigure[]{
\includegraphics[width=6.5 cm,height=6 cm]{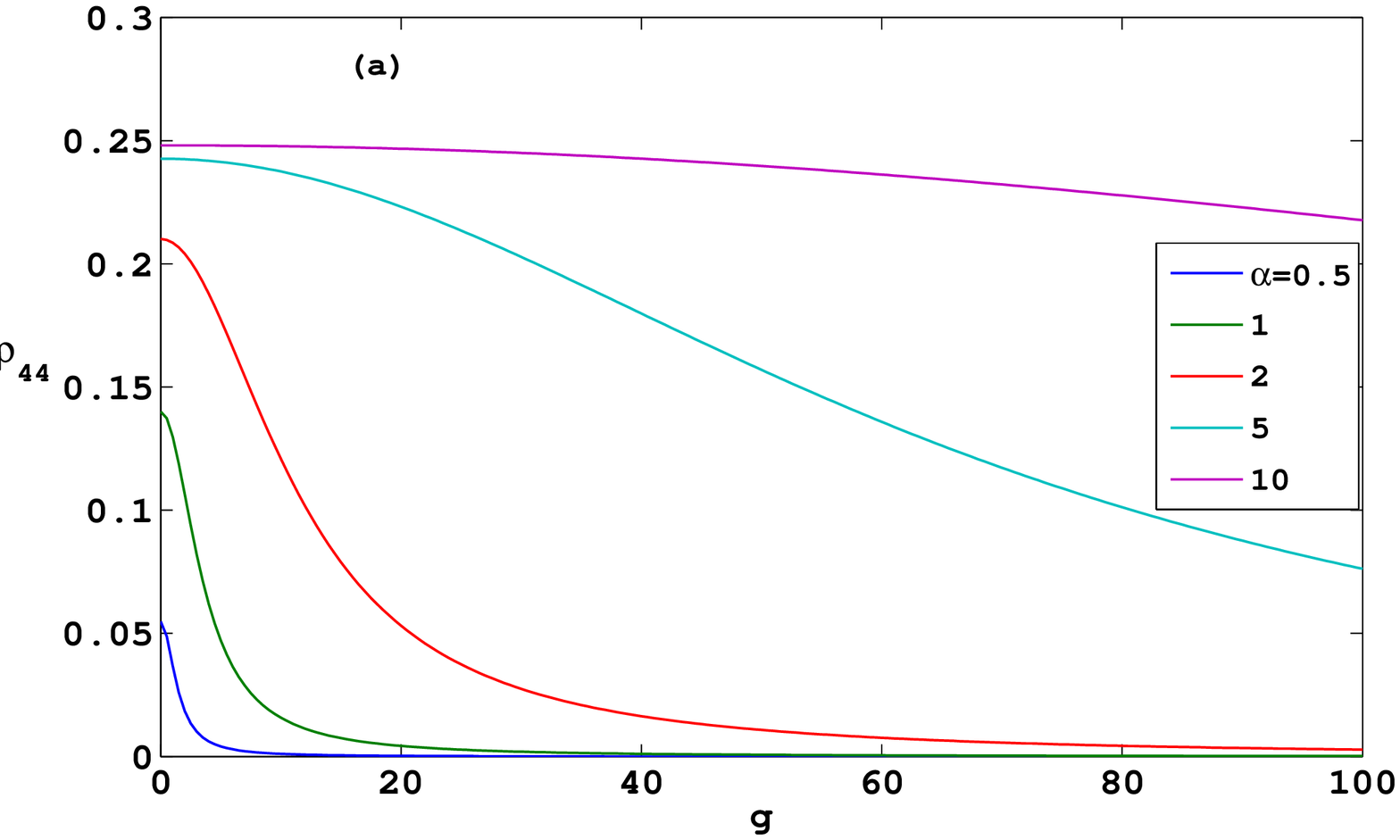}
\label{fig:Subfigure1}
}
\subfigure[]{
\includegraphics[width=6.5 cm,height=6 cm]{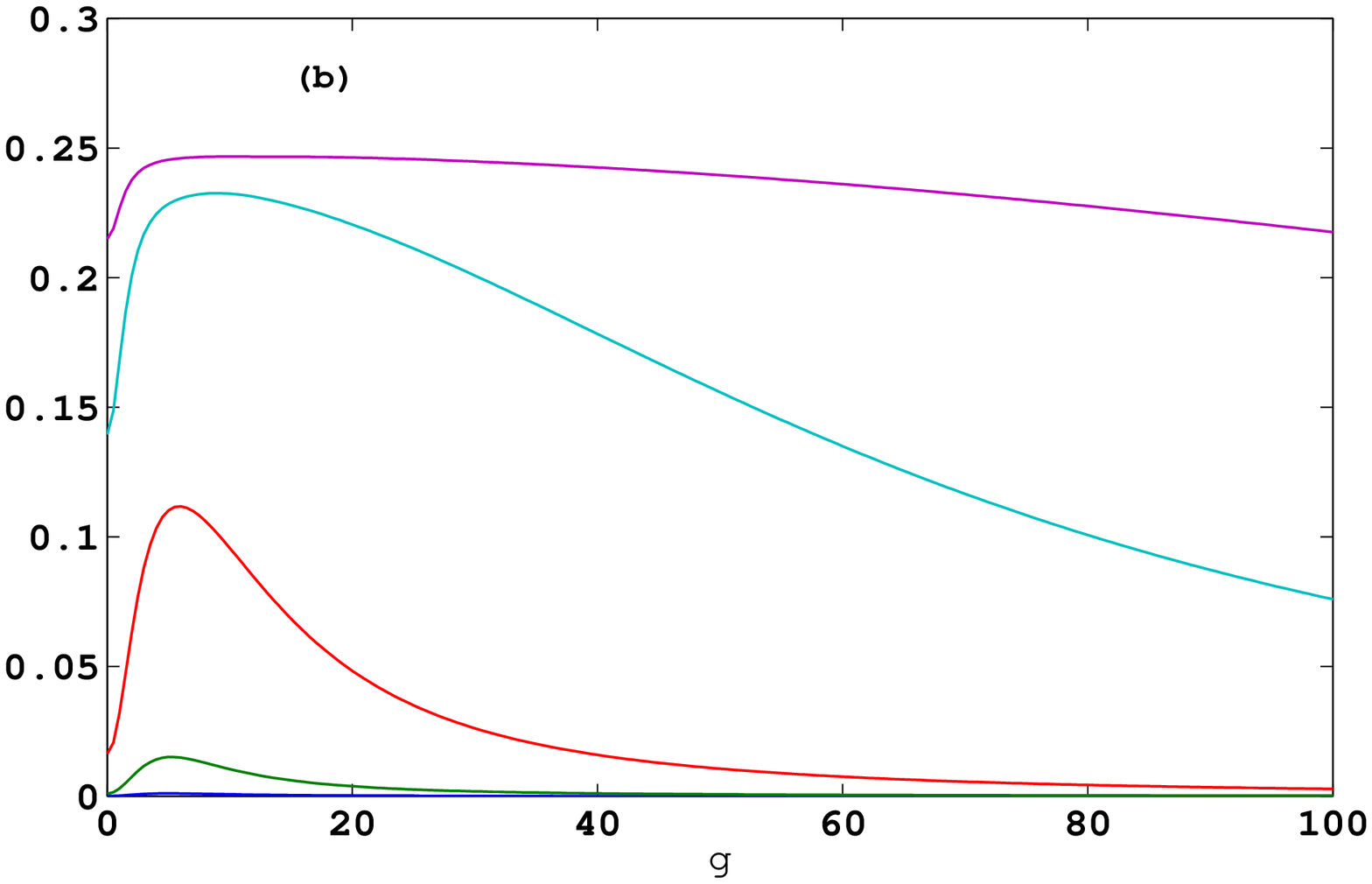}
\label{fig:Subfigure2}
}
\caption[Optional caption for list of figures]{\textit{Level populations $\rho_{44}$ for}  \subref{fig:Subfigure1} identical atoms  \subref{fig:Subfigure2} non-identical atoms}

\label{fig:Five}
\end{figure}

  Figures \ref{fig:Four} and \ref{fig:Five} elucidate the above mentioned behaviour, specifically of the level populations,  more clearly.  In Figure \ref{fig:Four}, the population of the two-atom excited state $\rho_{44}$ is plotted for both identical (a) and non-identical (b) atoms for a value of the Rabi field strength $\alpha=2$ and for a range of values of the dipole coupling strength as indicated in the figure.  There is a gradual decrease in the height of the central peak with increase in $g$ for identical atoms.  When it comes to the non-identical atoms, this behaviour is not repeated and it is observed that the decrease in peak height with increase in dipole coupling strength is non monotonic which strongly suggests a competition between $\alpha$ and $g$.  To understand this behaviour for a continuous range of dipole-coupling strength, we have studied the on-resonance behaviour of $\rho_{44}$ (the behaviour of the central peak) as a function of the dipole coupling parameter for different values of the Rabi field strength $g$.  The results are presented in Figure \ref{fig:Five}, where the on-resonance population of the two-atom excited state $\rho_{44}$ is plotted for different values of $\alpha$ as the dipole coupling strength is scanned from 0 to 100.


 It is observed from figure \ref{fig:Subfigure2} that for different values of $\alpha$, there exists a particular value of $g$ for which $\rho_{44}$ attains its maximum and then it begins to decrease.  This is not the case with identical atoms, wherein $\rho_{44}$ monotonically decreases with increase in $g$ as evident from  figure \ref{fig:Subfigure1}.  The non-monotonicity of the peak height  is seen to occur only for non-identical atoms. 
 
\begin{figure}[h!]

\centering
\subfigure[]{
\includegraphics[width=6.5 cm,height=5.5 cm]{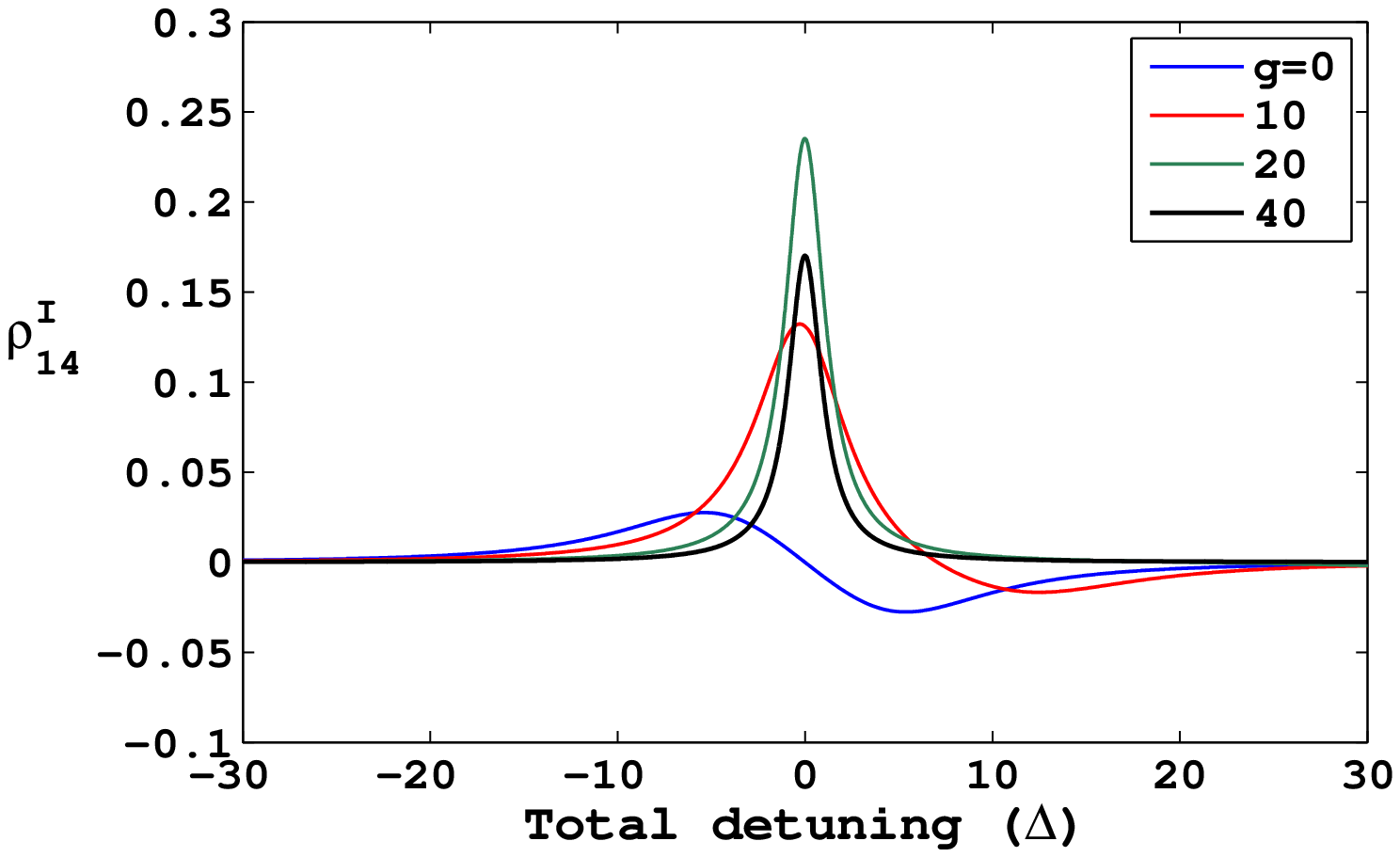}

\label{fig:SubFigure1}
}
\subfigure[]{
\includegraphics[width=6.5 cm,height=5.5 cm]{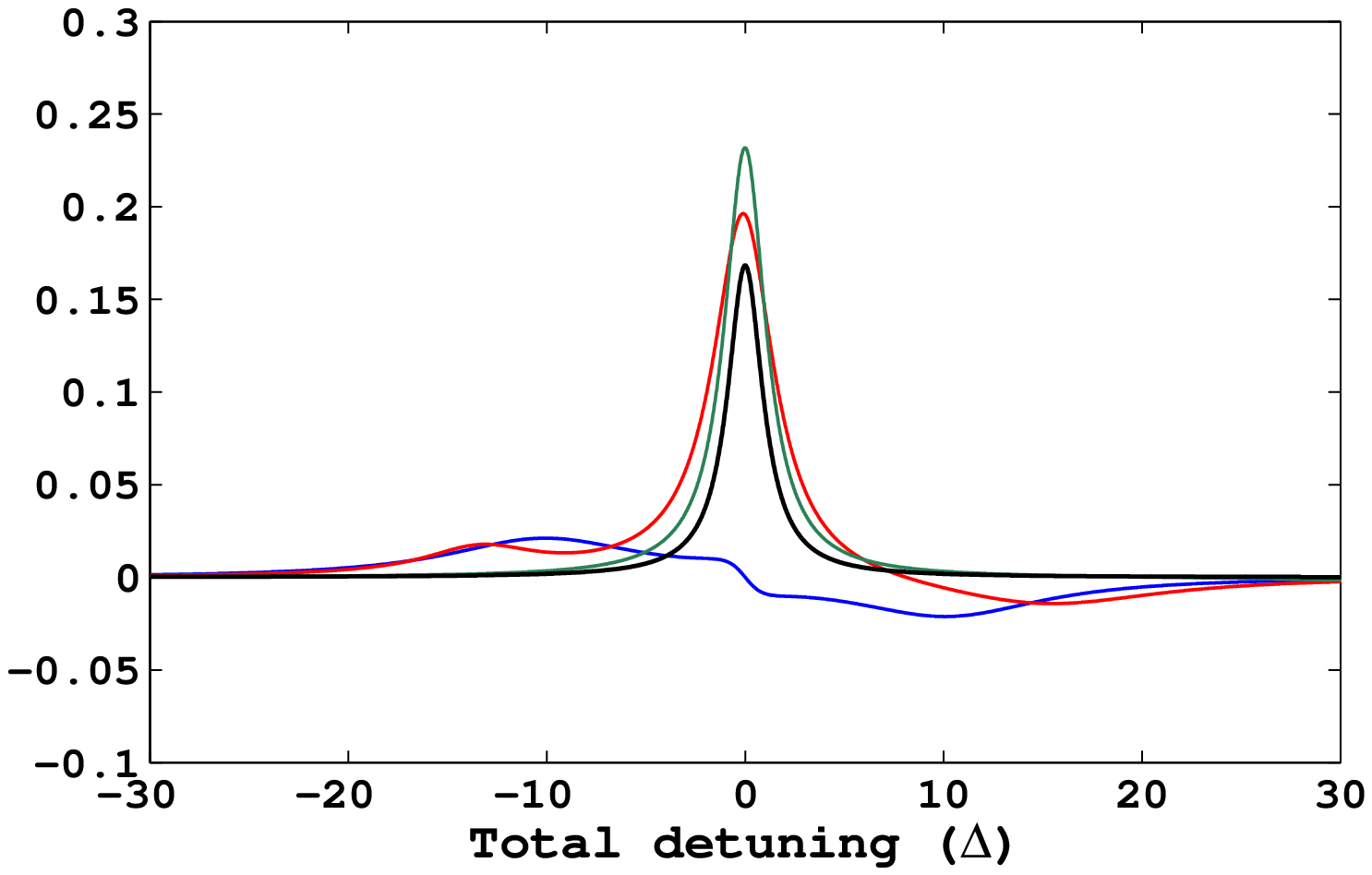}
\label{fig:SubFigure2}
}
\caption[Optional caption for list of figures]{\textit{Im $\rho_{14}$ for}  \subref{fig:SubFigure1} identical atoms  \subref{fig:SubFigure2} non-identical atoms}

\label{fig:Six}
\end{figure}

 The behaviour of the two-atom two-photon coherence Im $\rho_{14}$ is shown, both for identical  as well as non-identical  atoms  in \ref{fig:SubFigure1}  and  \ref{fig:SubFigure2} respectively.
 
The imaginary part of the atomic coherence $ \rho_{14} $ [cf. figure \ref{fig:Six}] seems to exhibit non monotonic behaviour 
 with respect to increase in $ g $ for both the cases.  The height of the central peak is found to increase up to a value of $ g =20$ and a further increase in $ g (=40$) results in a decrease of the same.  For the choice of detunings that are  chosen in this figure, the side bands are not well resolved.  However, for large enough detunings, we did observe that these side bands are indeed well resolved.
 \begin{figure}[h!]
\vspace{-10pt}
\centering
\subfigure[]{
\includegraphics[width=6.5 cm,height=6 cm]{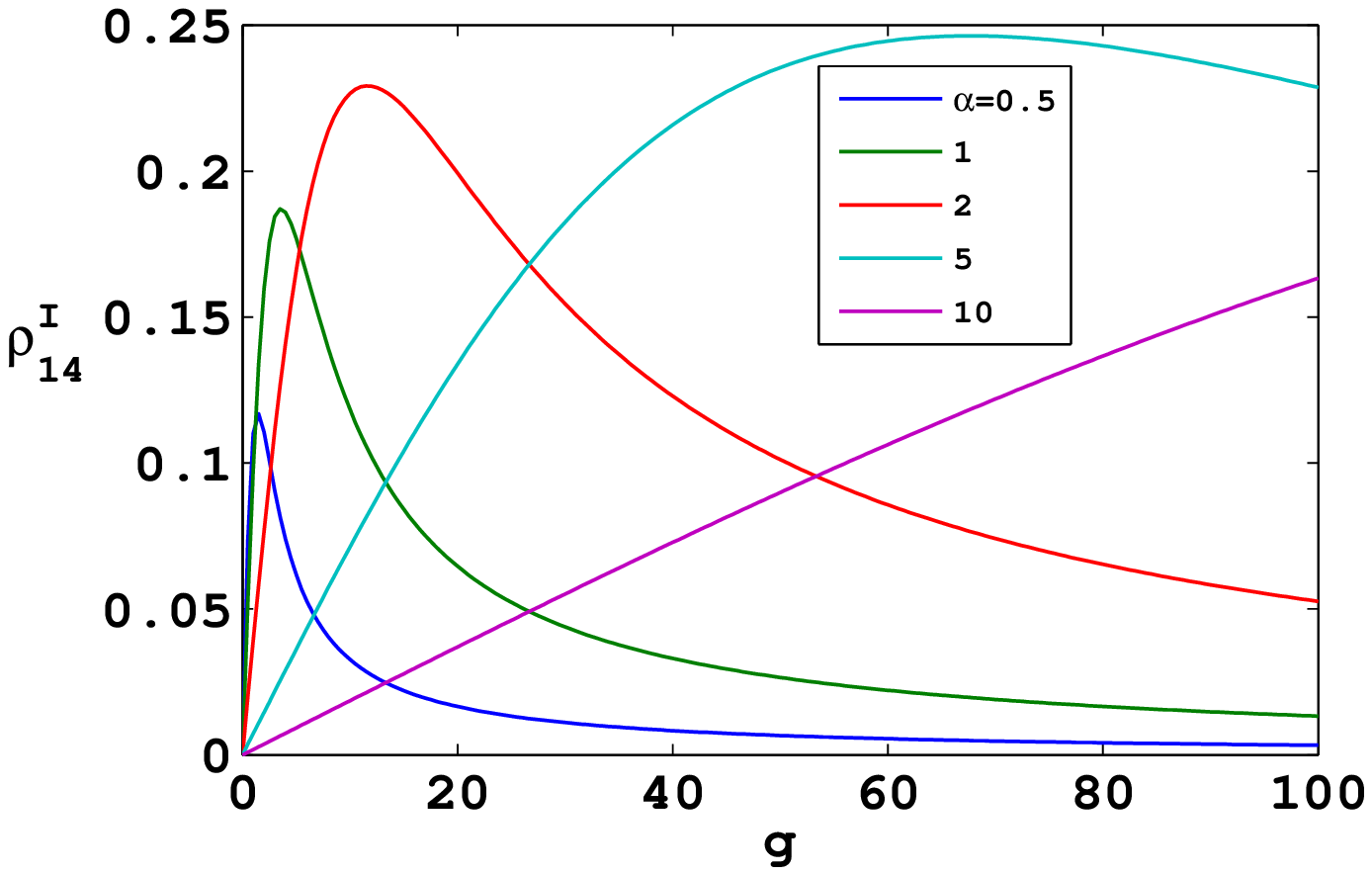}
%\vspace{-10pt}
\label{fig:SubFig1}
}
\subfigure[]{
\includegraphics[width=6.5 cm,height=6 cm]{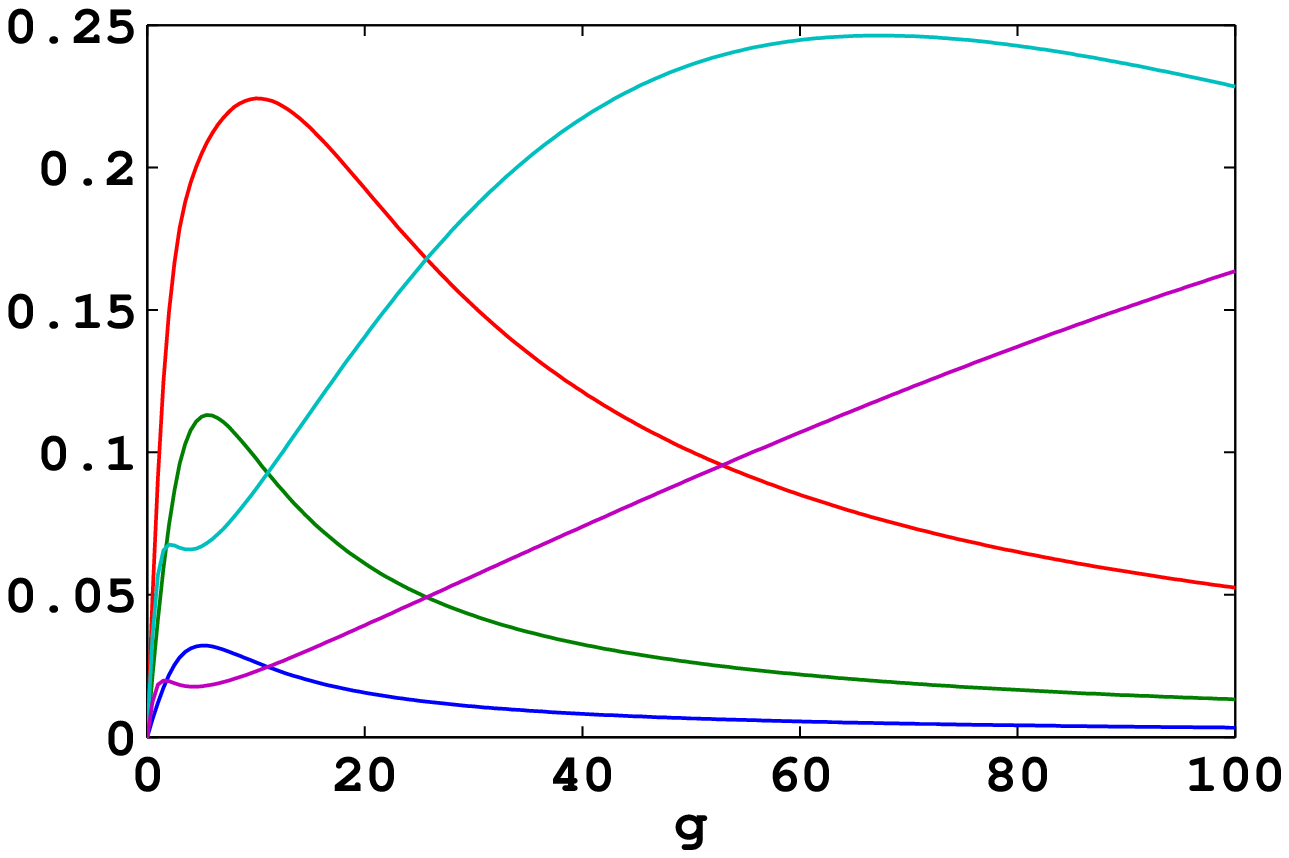}
%\vspace{-10pt}
\label{fig:SubFig2}
}
\caption[Optional caption for list of figures]{\textit{ The atomic coherence Im $\rho_{14}$ on resonance for}  \subref{fig:SubFig1} identical atoms  \subref{fig:SubFig2} non-identical atoms}
%\vspace{10pt}
\label{fig:Seven}
\end{figure}

In the results shown above, the atomic cooperative behaviour is incorporated through the dipole-dipole coupling strength $g$.  In the work of Varada and Agarwal \cite{r14}, they have included both the cooperative decay parameter as well as the dipole-dipole coupling strength.  We have carried out the same calculation and found that cooperative decay parameter acts like an additive feature to the dipole-dipole interaction strength.  In all the cases studied above, it is seen that the inclusion of the cooperative decay  causes faster decay from level $|\textbf{4}\rangle$ resulting in enhancing the effect of $g$.  But the features of non-monotonic behaviour of the level populations, coherences  that are reported above remain unaffected due to the inclusion of the cooperative decay parameter.

\section{CONCLUSION}
We have studied the effect of dipole-dipole interaction on a pair of two-level atoms interacting with a laser field.  The coupling of each of the atoms to the external radiation field, which is treated in the electric dipole approximation, and the strength of dipole interaction are incorporated phenomenologically by the parameters $\alpha$ and $ g $ respectively.   The level populations and the induced atomic coherence are studied for a wide range of parameters both for identical as well as non identical atoms.  The non-identical nature of the atoms is incorporated through different energy level separation while the other parameters such as the decay constants and coupling to the external field are assumed to be the same for both atoms, though in reality they may differ.  Extensive studies are carried out for the above mentioned features of the atomic system for a wide range of parameters with particular emphasis on the competing effects of dipole coupling parameter and the Rabi field strength.  In particular, the two photon resonance  which appears in the presence of the dipole coupling strength shows an increase upto certain values of the coupling strength and beyond the critical value, it decreases thus exhibiting a non-monotonic behaviour (only for the case of non-identical atoms). However, non monotonicity is seen in  the coherence $\rho_{14}$  both for identical as well as non-identical atoms.   The fluorescence  signal   shows qualitative agreement with the experimental results of ref \cite{r25}.

%\section{ACKNOWLEDGEMENTS}
Shaik Ahmed gratefully acknowledges the financial support from University Grants Commission, Govt. of India.
%\section*{References}

\end{document}